\documentclass[pdflatex,sn-mathphys-num]{sn-jnl}
\usepackage{graphicx}%
\usepackage{multirow}%
\usepackage{enumitem}
\usepackage{amsmath,amssymb,amsfonts}%
\usepackage{amsthm}%
\usepackage{mathrsfs}%
\usepackage[title]{appendix}%
\usepackage{xcolor}%
\usepackage{textcomp}%
\usepackage{manyfoot}%
\usepackage{booktabs}%
\usepackage{algorithm}%
\usepackage{algorithmicx}%
\usepackage{algpseudocode}%
\usepackage{listings}%
\usepackage{listings}
\usepackage{color}
\usepackage{xcolor}
\usepackage{float}
\usepackage{geometry}
\usepackage[justification=centering]{caption}
\usepackage{geometry}
\newgeometry{
tmargin = 0.75in , bmargin = 0.75in , lmargin = 0.75in , rmargin = 0.75in
}
\begin{document}    

\title[Classical RL Advantage over Quantum RL]{Hybrid-Quantum Neural Architecture Search for The Proximal Policy Optimization Algorithm}

\author[1]{\fnm{Moustafa} \sur{Zada}}\email{moustafa7zada@proton.me}

\affil[1]{\orgname{The University of Aleppo}, \state{Aleppo}, \country{Syria}}

\abstract{Recent studies in quantum machine learning advocated the use of hybrid models to assist with the limitations of the currently existing Noisy Intermediate Scale Quantum (NISQ) devices, but what was missing from most of them was the explanations and interpretations of the choices that were made to pick those exact architectures and the
differentiation between good and bad hybrid architectures, this research attempts to tackle that gap in the literature by using the Regularized Evolution algorithm to search for the optimal hybrid classical-quantum architecture for the Proximal Policy Optimization (PPO) algorithm, a well-known reinforcement learning algorithm, ultimately the classical models dominated the leaderboard with the best hybrid model coming in eleventh place among all \textbf{unique} models, while we also try to explain the factors that contributed to such results,and for some models to behave better than others in hope to grasp a better intuition about what we should consider good practices for designing an efficient hybrid architecture.}


\maketitle
\section{Introduction}

Quantum machine learning (QML) has emerged in recent years as a promising approach to advance classical machine learning, considering that they can evaluate classically intractable functions \cite{killoran_hybrid_nodate} since they have properties that the classical computational models don't have access to like the exponential Hilbert space, entanglement, and the parallelistic nature of quantum computation in the presence of superposition.

But, Since we are still in the NISQ era, with limited quantum computers, many studies have discussed the use of hybrid architectures in the hope that they can get an advantage from the currently existing quantum computers or if they could find a way to distribute the work harmonically and efficiently between the classical and the quantum part, each part with what they are good at.
An important Distinguish that had to be made here is that when we are not considering variational quantum circuits as hybrid classical-quantum models, since they use the classical computer only to train and optimize the quantum circuit but they don't interfere with the actual learning and the inference of the model, instead, we treat the VQC itself as a layer, a quantum layer, in the neural network of the hybrid model along with classical layers with artificial neurons.

In this paper, we apply the proximal policy optimization (PPO) algorithm to the classical CartPole environment problem to ascertain whether the quantum-enhanced PPO algorithm gives any advantage over the classical version. In particular, we use a genetic algorithm-esque version of quantum PPO to train the system. 

This paper is divided as follows:

In Sec.\ref{related-work}, we examine analogous studies to the experiments we have performed in order to gauge what has been done, and how we can improve upon it.

In Sec.\ref{background}, we provide a brief synopsis of the operation of the PPO algorithm. 

In Sec.\ref{experiments}, we provide the details of the experimental constraints considered in the training and the results from the training.

In Sec.\ref{conclusions} and \ref{future}, we reflect on the results obtained and allude to future considerations we hope to investigate.

\section{Related Work}\label{related-work}\
As demonstrated in this comprehensive survey on Quantum Reinforcement Learning (QRL), most of the literature focuses on developing the VQC themselves, their ansatz (or architecture often called in the literature , contrary to what we mean by architecture in this paper), and using them as function approximation instead of the classical neural networks, but some extensions are made to hybrid quantum classical models like those we are working on in this paper, for instance, Chen(2023)  \cite{chen_asynchronous_2023} constructed a hybrid model for the A3C actor-critic algorithm that has a linear layer before the VQC, which was a well known and widely used ansatz for VQC, and it shown comparable or even superior results to other classical models with comparable sizes, in another work by Chen et al.(2023)\cite{chen_deep-q_2023} they used also a hybrid model with two convolutional layers, a linear layer, a quantum layer, then a linear layer to solve a maze navigation problem using Deep Q-Learning algorithm, the model demonstrated comparable although slightly worse performance than the classical counterpart but with less parameters, a well-known benefit from using VQC.\\
Another interesting paper by Lockwood and Si(2021)\cite{pmlr-v148-lockwood21a} that tried to use QRL for a more challenging task, playing atari games, the authors experimented with 12 different hybrid configurations while all of them had a common feature, having a classical layer before the quantum layer to reduce the high dimensions of the data coming from the atari game to the VQC, which as the authors suspect, the extreme reduction of size lead to all models failing to learn and the classical model with sutabele number of parameters performing much better, they came to this conclusion considering that \textit{a}. the models don't have anything fundamentally wrong with them or bugged code since they can learn other simpler tasks, \textit{b}. a classical model with $O(1o^4)$ doesn't learn either, indicating that the problem is too hard for small models quantum or classical to learn the underlying function approximations. 
a notable work by Dragan et al.(2022)(\cite{dragan_quantum_2022}
have they conducted experiments with 19 different hand-picked and well-known ansatz for VQC as a part of a hybrid model for a quantum variation of PPO that has a VQC then a 4-neuron linear layer as the architecture for both networks(actor and critic), in this study the authors to analyze the results that they got in hope to understand why some VQC ansatzes are performing better than others while having the same 4-neuron pre-processing layer with same hyperparameters, three metrics were considered, expressibility, entanglement capability, and the effective
dimension, eventually, No direct and clear correlations were shown to the performance, keeping this question unanswered.\\
we build on that work to search without \textbf{Human Bias} for more than 1000 iterations for PPO since most of the work done above either hand-picked famous ansatz architectures or just stated that the model that they have picked proved to be the most promising after conducting experiments.
Thus, this work investigates this gap in the literature, using an evolutionary algorithm for Neural Architecture Search (NAS) called Regularized Evolution \cite{real_regularized_2019} which proved to be quite effective in searching for an optimal or sub-optimal architecture surpassing currently used reinforcement learning approaches for NAS, we used this algorithm to search for an efficient architecture-finding the optimal one is not guaranteed-
for the Proximal Policy Optimization Algorithm \cite{schulman_proximal_2017}.
This work also tries to explain the experimental results obtained by running the search algorithm for \textbf{1000+ iterations}.

\section{Background}\label{background}
\subsection{Proximal Policy Optimization}
Briefly, \textbf{PPO} is an actor-critic algorithm, meaning there are two models involved, an Actor which \textit{Acts} in the environment and takes actions, and a Critic which evaluates the Actor's actions contributing to the objective function(the loss function) of the Actor by determining the goodness/badness of the actions at a specific observation state.
generally speaking, the only interesting thing about PPO is the brilliantly crafted loss function, which is clipped to ensure that the changes of the network parameters in the weights space are small and do not diverge widely if you encounter an action state pair that the model thinks is too good or too bad, taking moderate steps in such extreme cases makes the training much more stable. 
the objective function for the Actor without clipping: 
\begin{equation}
L^{CPI}(\theta) = \hat{\mathbb{E}}_t \left[ \frac{\pi_\theta(a_t \mid s_t)}{\pi_{\theta_{\text{old}}}(a_t \mid s_t)} \hat{A}_t \right] 
= \hat{\mathbb{E}}_t \left[ r_t(\theta) \hat{A}_t \right].
\end{equation}
After clipping: 
\begin{equation}
L^{CLIP}(\theta) = \hat{\mathbb{E}}_t \left[ 
\min \big( r_t(\theta) \hat{A}_t, 
\text{clip}\big(r_t(\theta), 1 - \epsilon, 1 + \epsilon\big) \hat{A}_t \big) 
\right].
\end{equation}
the complete loss function for the whole model, consisting of the loss for the Actor, the loss for the Critic, and the entropy of the Actor's actions probabilities which introduces some randomness to our model to encourage exploration rather than consistent exploitation: 
\begin{equation}
L_t^{CLIP+VF+S}(\theta) = \hat{\mathbb{E}}_t \left[ 
L_t^{CLIP}(\theta) - c_1 L_t^{VF}(\theta) + c_2 S[\pi_\theta](s_t) 
\right].
\end{equation}
Note that our experiment used the same architecture for both networks, inspired by famous, well-documented implementations of the PPO algorithm \cite{shengyi2022the37implementation}.
\subsection{Regularized Evolution}
The way that Regularized Evolution\cite{real_regularized_2019} works is quite straightforward: we start with a \textbf{P} number of randomly generated architectures, and we keep only \textbf{P} number of architecture at our hand at any time, we call it the \textbf{population}, and we keep another list of arch. we call the \textbf{history} were we store every architecture we encounter, then we iteratively and randomly pick a sample of size \textbf{S} from the population and mutate the highest scoring one of them, then, train it, add the newly trained model to the population, and delete the oldest arch. in the population even if it has the highest score, statistically its guaranteed to get back to that arch. or to another even better one so it's safe to do this, and that's why the authors called it aging evolution throughout the paper.

\section{Experiments}\label{experiments}
When designing the mutation, you essentially create the search space for the search algorithm, meaning that the more that you add freedom to the mutations that could be taken, the less human bias that you introduce into the search, and the higher the chance to find interesting and unusual results, the mutations act on a DNA sequence that has a detailed explanation in the code repository \cite{zada_notitle_nodate}.

\subsection{The Mutations Used}\label{used mutations}
Note that some tuning to the neighboring layers is required after applying most of these mutations. The details of such changes are present in the code \cite{zada_notitle_nodate}.
\begin{enumerate}
\item \textbf{Add a Classical Layer:} Add a classical layer at a random position.
\item \textbf{Remove a Classical Layer:} Remove a Classical Layer picked randomly.
\item \textbf{Add a Quantum Layer:} Add a quantum layer in a location chosen at random.
\item \textbf{Remove a Quantum Layer:} If a quantum layer exists, remove it.
\item \textbf{Alter a Layer [Add]:} Add a random (constrained) number of neurons or qubits to a randomly chosen layer.
\item \textbf{Alter a Layer [Remove]:} Remove a random (constrained) number of neurons or qubits to a randomly chosen layer.
\item \textbf{Change the repetitions of the ansatz:} Pick a quantum layer at random and change its number of layers in the ansatz.
\item \textbf{Change the Entanglement type of the Ansatz:} Pick a quantum layer at random and change its entanglement type.
\item \textbf{Change the Activation Function of a Layer:} Pick a classical layer randomly and change its activation function. The supported functions are either Tanh or ReLU. 
\item \textbf{Identity:} Do not change anything.
\end{enumerate}

\subsection{Experimental Constraints}
\begin{enumerate}
\item Maximum number of layers = 10. 
\item Maximum number of neurons in a classical layer = 64.
\item Maximum number of qubits in a quantum Layer = 10.
\item Minimum number of neurons in a classical Layer = 2.
\item Minimum number of qubits in a quantum layer = 2.\\
\end{enumerate}
\textbf{An Important Detail} of this implementation is that there are two kinds of quantum layers used, layers that output only the expectation value of at most $n$ qubits, where $n$ being the number of qubits of the layer under consideration, and layers that outputs $2^n$ measurement results for all the possible bit strings using the default shots number of 1024, the former was used if there are two quantum layers adjacent since the input of any quantum layer should be equal to its qubit count $n$ thus using the second option was not plausible, or the network would get very shallow if two quantum layers come after each other because of a random mutation, cause in that case we should use logarithmic values of the second quantum layer's qubits $n_2$ as the qubit number of the first quantum layer $n_1$.
\[\log_2(n_2) = n_1\]
The latter approach was used if there is a classical linear layer after that quantum layer, which takes any number of inputs.\\
This idea is not very usual. It was taken to limit the extreme shallowness of the network in case two quantum layers come together and to limit the human bias to allow the search algorithm to be as free as possible. However, it might have negatively affected the results, which is unclear.\\
The Experiment was conducted using PyTorch for the Classical Part of the models, and the training, using the ADAM optimizer for the Quantum Layers Pennylane \cite{bergholm_pennylane:_2022}, was used. The experiment took approximately 20 Hours. Having the Classical Layers with the parameters of the quantum layers on the GPU -- an RTX 3070 with 8GB of VRAM -- and the quantum layer execution on the CPU, this configuration might seem un-optimized, but this was the best configuration possible considering missing features in Pennylane and the small width of our circuits makes the execution on the GPU not so intriguing.
The used environment is \textbf{CartPole-V1}, a famous OpenAI gym 
environment \cite{brockman2016openai}.

\subsection{Results} 
Out of all the 1000+ iterations, \textbf{666} unique models were found.
\textbf{390} models were purely classical without any quantum layers, \textbf{236} models had one quantum layer, \textbf{34} models had two quantum layers, \textbf{4} had three layers, and \textbf{2} had four layers, without having any higher number of quantum layers in any model. 
The Average score of the model over 10 different episodes from 0 to 500 will be used as the sorting factor between all the 1000+ models, the highest scoring model being a classical model with a score of 442.6/500.\\

\begin{figure}[!htbp!]
    \centering
    \includegraphics[width= 500pt , height = 150pt]{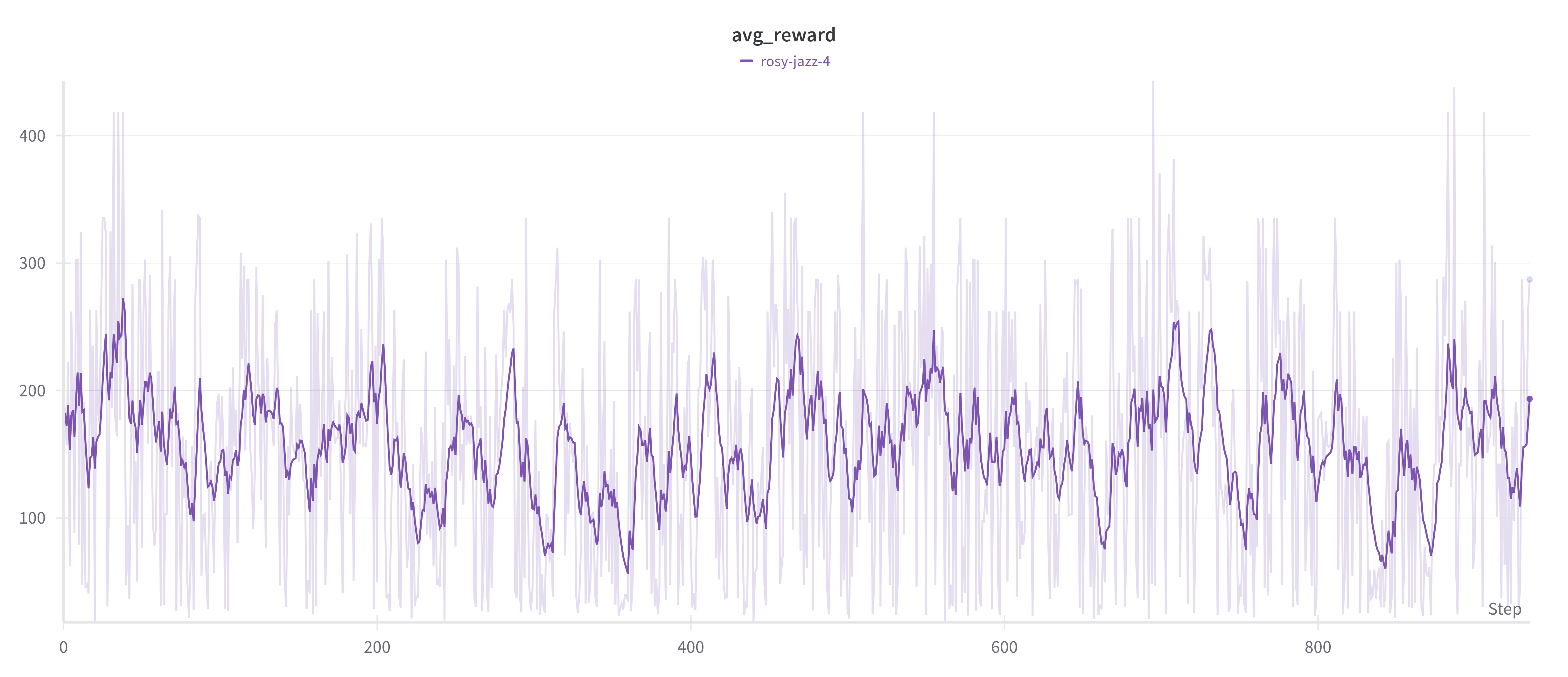}
    \caption{The Average Reward on The Span of approx. 950 Iterations}
\end{figure}

If we discard the duplicated models(since when we checked if the model is already trained to use existing values from a previously trained instance of that model)
then we would have a leaderboard of the top 10 models shown below in Tab.\ref{tab:my_label}, you can see that the first 8 models are Classical models and the 11th unique model is a hybrid model with a small quantum layer in it with two Qubits only.
Note that all 11 models achieved at least one perfect score of 500 in one of the evaluating episodes, with the 9th one having three perfect episodes and one episode score equal to only 17 points! , which made it lose to other consistent classical models.
This is rather an interesting and unusual behavior that the hybrid model has shown.\\

\begin{table}[!h!]
\centering
\begin{tabular}{|l|c|}
\hline    
DNA Sequence &Average Reward\\
\hline
C 37, T,    C 41, T, C 2, R, C 24, T, C 5, T, C, 1	& 442.6  \\
C 50, T, C 13, R, C 2, T, C 9, R, C 64, T, C 42, T, C 38, R, C 1	
& 437.7  \\
C 50, T, C 13, R, C 2, T, C 9, R, C 64, T, C 42, R, C 38, R, C 1	& 418.6  \\
C 50, T, C 16, R, C 2, T, C 50, R, C 64, T, C 42, R, C 38, R, C 1 & 389.7 \\ 
C 57, R, C 41, T, C 2, R, C 24, T, C 5, T, C, 1	& 381.2  \\ 
C 50, T, C 13, R, C 2, T, C 59, R, C 55, T, C 42, R, C 38, R, C 1	& 370.6  \\
C 50, T, C 16, R, C 3, T, C  9, R, C 64, T, C 42, R, C 38, R, C 1 & 366.9 \\  
C 50, T, C 13, T, C 2, T, C 9, R, C 64, T, C 42, R, C 38, R, C 1 & 	355.1  \\
C 51, R, C 24, T, C 5, T, C, 1 &	341.5  \\
C 37, R, C 38, T, C 6, T, C 5, T, C 1 & 340.8 \\ 
C 50, T, C 13, R, C 2, T,\textbf{Q 2 F 3}, C 9, R, C 55, T, C 29, R, C 42, R, C 38, R, C 1& \textbf{339.7}  \\
\hline
\end{tabular}
\caption{Results from the 11 best-performing experiments.}
\label{tab:my_label}
\end{table}
The Used encoding method is angle encoding, and in the shown DNA sequences, \textbf{C} stands for a classical layer with the number of its neurons next to it, \textbf{R} and \textbf{T} are activation functions and \textbf{Q} stands for a quantum layer with its number of Qubits next to it, along with the entanglement type of the layer, either \textbf{L} for \textit{BasicEntanglingLayer} or \textbf{F} for \textit{StrongEntanglingLayer} with the number of repetitions next to it. 

\section{Conclusions}\label{conclusions}
Empirically, we can conclude some observations and patterns in the data:
\begin{itemize}

    \item One can see in Table \ref{tab:my_label} that the training is not stable, while most of the instability comes from tweaking the quantum layers, the smallness of the networks which makes them so sensitive to any change in the network is also a big contributor to the instability, which means that the used mutations are too free and random for such an experiment. 

    \item The Idea of using a big quantum layer that we do not measure all of its qubits, and relaying on the entanglement to propagate the information between the qubits(throughout the width of the circuit) is not a great idea, as it is showing in the data\cite{zada_notitle_nodate} most of the lowest performing models share that feature, which hints that having a healthy backpropagation throughout your model, where all parameters do make a difference to the loss value,is far better than this case where the optimizer tunes the parameters without a real difference to the loss because the learning is dependent on the propagation of information via entanglement rather than directly.
    
    \item Quantum Layers with 4 or 3 Layers were only present in the random generation of the population phase of the evolution algorithm and the algorithm didn't favor them again due to their bad performance.  
    \item Quantum layers with many qubits proved to be harder to train using the used hyper-parameters, which made them not favorable by the search algorithm.
    
    \item By mutating the best hybrid-model \ref{tab:my_label} by hand, we could see that changing the entanglement from \textit{Strong} to \textit{Basic} results in the average reward being halved(while CartPole is not the most accurate environment which may result in the huge variation that happens) this suggests that having a quantum layer with a small number of qubits and high entanglement is favorable. 
    
    \item Finally, this experiment showed that quantum variational circuits as quantum layers in hybrid models in their current form do not hint at any advantage over using well-designed classical models. 
     
\end{itemize}

\section{Future Work}\label{future}
In coming versions of this exact paper or continuations of this paper, I hope that I can test more environments to get more coherent data and provide stronger evidence for the conclusions, an effort that I have not made yet due to the computational demand that other environments require since the used one is considered quite simple, and it took 20 hours of training despite that. In addition, more neural architecture search algorithms should be explored. Ideally, \textbf{new custom} hybrid Quantum-Classical NAS Algorithms should be developed to aid the existing efforts to adopt hybrid models in real-world applications.

\section*{Acknowledgments}
I would like to thank Muhammad Al-Zafar Khan for his insights, mentoring, and help. Additionally, this research benefited from insights from the literature review on the search for neural architectures by C. White et al. \cite{white_neural_2023} 
on a final note, it came to my attention that the \textit{ADD a Quantum Layer}\ref{used mutations} is bugged in an edge case, but I don't think that It affected the training in a concerning way.\\
\noindent\textbf{Disclaimer:} This research was neither funded nor supervised by any organization or institution, including the home university of the author. It represents an independent effort by the author.

\section*{Code and Data Availability}
The code and data for this project may be accessed from the author's GitHub account \cite{zada_notitle_nodate}.

\bibliography{citations}


\end{document}